\documentclass[onecolumn,showpacs,amsfonts,aps,prc,nofootinbib,floatfix,%
superscriptaddress]{revtex4}

\usepackage{amsmath}
\usepackage{bm}
\usepackage{graphicx}

\usepackage{epsfig}
\newcommand{\beq}{\begin{equation}}
\newcommand{\eeq}{\end{equation}}
\newcommand{\bea}{\vspace{0.25cm}\begin{eqnarray}}
\newcommand{\eea}{\end{eqnarray}}

\newcommand{\kb}{{{\bf k}}}

\newcommand{\bb}{{{\bf b}}}

\def\lsim{\mathrel{\rlap{\lower4pt\hbox{\hskip1pt$\sim$}}
    \raise1pt\hbox{$<$}}}         
\def\gsim{\mathrel{\rlap{\lower4pt\hbox{\hskip1pt$\sim$}}
    \raise1pt\hbox{$>$}}}         

\newcommand{\landau}{L.D.~Landau Institute for Theoretical Physics,
        GSP-1, 117940, Kosygina Str. 2, 117334 Moscow, Russia}

\begin{document}


\title{
Effect of meson cloud on 
the jet nuclear modification factor in $pA$ 
collisions
}
\date{\today}

\author{B.G.~Zakharov}\affiliation{\landau}

\begin{abstract}
We study the effect of the nucleon meson cloud on
centrality dependence of the jet nuclear modification factor $R_{pA}$.
We find that the meson-baryon Fock components  may lead to a 
noticeable deviation of $R_{pA}$ from unity.
Our results for $R_{pA}$ show 
the same tendency as that
observed by ATLAS in $p+Pb$ collisions at $\sqrt{s}=5.02$ TeV.
The meson cloud suppresses the central jet events and
enhances the peripheral jet events.
But quantitatively the effect is somewhat smaller than in the
data.
  
\end{abstract}
%

\maketitle

\section{Introduction}
Factorization of hard and soft process \cite{Collins1} suggests that
in the cross section of hard reactions the soft physics can be accumulated
in parton distribution functions (PDFs) of colliding particles.
However, the factorization theorems do not forbid the existence
of correlations between hard and soft final particles.
The correlations of this type have been observed in recent measurements 
by ATLAS \cite{ATLAS1} of the centrality 
dependence of the jet nuclear modification factor $R_{pA}$ for $p+Pb$
collisions at $\sqrt{s}=5.02$ TeV. The $R_{pA}$ for jet production 
is defined as
\beq
R_{pA}=\frac{dN_{pA}^{jet}/dp_{T}dy}
{N_{coll} dN_{pA}^{jet}/dp_{T}dy}\,,
\label{eq:10}
\eeq 
where $N_{pA}^{jet}$ and $N_{pp}^{jet}$ are the jet yields in $pA$ and $pp$
collisions, and $N_{coll}$ is the number of the binary 
collisions.
In \cite{ATLAS1} it has been observed that the jet nuclear modification factor
$R_{pPb}$ in the broad (minimum bias) $0-90$\% centrality region is 
close to unity.  However, it is not
the case for narrow centrality bins. For high $p_T$ the 
$R_{pPb}$ has been  found  to be suppressed in central events 
($R_{pPb}<1$) and to be enhanced in peripheral events ($R_{pPb}<1$).
The effect is more pronounced for the proton-going rapidities ($y>0$).
In principle the suppression of the jet $R_{pPb}$ in the central events
might arise from the final state interaction effects in the small-size
quark-gluon plasma, if it is formed in $pPb$ collisions \cite{Z_mQGP2}.
However, the fact the minimum bias jet $R_{pPb}$ observed in \cite{ATLAS1}
is approximately consistent with unity, says that the potential effect 
of the plasma mini-fireball is small (or it is well compensated by the
medium effects in $pp$ collisions  \cite{Z_mQGP1} due to modification
of the denominator of (\ref{eq:10}) \cite{Z_mQGP2}).  
In \cite{Skokov1} it was proposed that the ATLAS data \cite{ATLAS1}
can be explained by the initial state correlations of the hard and soft
partons in the wave function of the projectile proton. The mechanism
of \cite{Skokov1} assumes that in the presence of an energetic
parton (which is necessary for a hard process to occur) 
the number of soft partons in the projectile is suppressed.
Then, assuming that the multiplicity of soft particles produced in 
the underlying events (UEs) with jet production is proportional
to the number of soft partons in the projectile proton,
it leads naturally to correlation of the jet $R_{pA}$ with the 
multiplicity/centrality of the UEs. In the recent more sophisticated analyses
\cite{Armesto,Majumder} this parton level mechanism has been addressed 
within the Monte-Carlo generators PYTHIA and HIJING. But only a very 
crude agreement with the ATLAS data \cite{ATLAS1} has been attained. 
However, of course, similarly to \cite{Skokov1}, the analyses 
\cite{Armesto,Majumder} 
are of a qualitative nature.  Because, due to the nonperturbative physics 
of the UE, it impossible to obtain robust predictions for the 
multiplicity/centrality dependence of $R_{pA}$ within the parton level schemes.

The purpose of the present work is to study the effect of the 
meson-baryon Fock components in the proton on the jet nuclear modification
factor. It is known that the total weight of the meson-baryon Fock states 
in the fast physical nucleon may be as large as $\sim 40$\% \cite{ST}.
The meson cloud of the proton plays an important role in the flavor
dependence of nucleon PDFs in deep inelastic scattering (DIS),
and is probably responsible for the violation of 
the Gottfried sum rule \cite{ST}.
The emergence of the centrality dependence of $R_{pA}$ in the scenario 
with the meson cloud is conceptually very similar to the partonic 
mechanism of \cite{Skokov1}. In this scenario the hard process
selects in the projectile proton wave function fluctuations with a reduced
fraction of the meson-baryon states (as compared to the soft interactions).
It results in suppression of the UE multiplicity in jet production at high
$p_T$, that should lead to centrality dependence of $R_{pA}$ due
to difference in the centrality categorization for minimum bias (soft) events 
and jet events. In the present analysis we simulate the UE activity 
in jet events within the Monte-Carlo Glauber (MCG) wounded nucleon model
with the meson cloud developed in \cite{Z_MCGL1,Z_MCGL2}.
We find that a considerable part of the centrality dependence
of the jet nuclear modification factor measured by ATLAS \cite{ATLAS1} 
may be explained by the meson-baryon Fock components of the proton.  

\section{Theoretical framework}
Our treatment of the meson-baryon components is similar to that  
in previous analyses of the meson cloud effects in DIS 
based on the infinite momentum frame (IMF) picture
\cite{ST,Zoller_MB,Zoller_MB1,MST_MB}. In this picture the physical nucleon IMF 
wave function reads 
\beq
|N_{phys}\rangle=\sqrt{W_{N}}|N\rangle+
\sum_{MB}\int dxd\kb\Psi_{MB}(x,\kb)|MB\rangle\,,
\label{eq:20}
\eeq
where $N$, $B$, and $M$ denote the bare baryon and meson states, 
$x$ is the meson fractional longitudinal momentum,
$\kb$ is the tranverse meson momentum, $\Psi_{MB}$ is
the $MB$ probability amplitude, $W_N=1-W_{MB}$ is the weight of the 
one-body Fock state in the physical nucleon, and
\beq
W_{MB} =\sum_{MB}\int dxd\kb|\Psi_{MB}(x,\kb)|^2
\label{eq:30}
\eeq
is the total weight of the $MB$ Fock components. 
The proton PDF for a parton $i$, $D_{i/p}$, corresponding to the Fock 
state decomposition (\ref{eq:20}), can be written as \cite{ST}
\beq
D_{i/p}(x,Q^2)=W_N\tilde{D}_{i/p}(x,Q^2)
+\int_{x}^{1} \frac{dy}{y}\tilde{D}_{i/M}(x/y,Q^2)f_{M/p}(y)
+\int_{x}^{1} \frac{dy}{y}\tilde{D}_{i/B}(x/y,Q^2)f_{B/p}(y)\,.
\label{eq:40}
\eeq
Here $\tilde{D}_{i/M}$ and $\tilde{D}_{i/B}$ are the PDFs for the 
bare particles, and  $f_{M,B/p}$ are the $p\to M,B$ splitting functions
given by
\beq
f_{M/N}(x)=\int d\kb |\Psi_{MB}(x,\kb)|^2\,,
\label{eq:50}
\eeq
\beq
f_{B/N}(x)=\int d\kb |\Psi_{MB}(1-x,\kb)|^2\,.
\label{eq:60}
\eeq

The analyses of the meson effects in DIS \cite{ST,Zoller_MB,Zoller_MB1,MST_MB}
show that in the Fock state decomposition (\ref{eq:20}) it is enough to include
$\pi N$, $\pi\Delta$, $\rho N$ and $\rho \Delta$ two-body systems.
The total weight of these states in the physical nucleon is
about $40$\% \cite{ST} with the dominating contribution
from the $\pi N$ states. For simplicity, we neglect the difference in the PDFs 
generated by the above four two-body states, and treat them as one
effective $\pi N$ state with normalization $W_{MB}=0.4$.
This is a reasonable assumption, because the $\Delta$ and $\rho$ PDFs should
be close to that for $N$ and $\pi$. Following the analyses of DIS 
\cite{ST,MST_MB} we evaluated 
the $f_{\pi/p}$ splitting function using the ordinary $\gamma_5$ pion-nucleon
vertex with the dipole formfactor 
\beq
F=\left(\frac{\Lambda^2+m_N^2}{\Lambda^2+M^{2}_{\pi N}(x,\kb)}\right)^2\,
\label{eq:70}
\eeq
with $\Lambda=1.3$ GeV (see \cite{Z_MCGL2} for details), $M_{\pi N}$
is the invariant mass of the $\pi N$ state in the IMF. 

We write the jet cross section ($\sigma(p_T,y)=d\sigma/dp_Tdy$) as a sum
\beq
\sigma(p_T,y)=\sigma_N(p_T,y)+\sigma_{MB}(p_T,y)\,,
\label{eq:80}
\eeq
where the first term on the right-hand-side of (\ref{eq:80})
corresponds to the one-body contribution to the proton PDFs from the first
term in (\ref{eq:40}), and the second term describes the effect 
from the last two terms in (\ref{eq:40})  due to the two-body Fock components.
In jet events the dynamics of the UEs in $pA$ collisions depends crucially 
on the relative contribution of the $\sigma_N$ and $\sigma_{MB}$ 
to the total jet cross section, because it controls the probabilities
of the $N$ and $MB$ states in the hard process given by
\beq
W_N^j(p_T,y)=\frac{\sigma_N(p_T,y)}{\sigma_N(p_T,y)+\sigma_{MB}(p_T,y)}\,,
\label{eq:90}
\eeq
\beq
W_{MB}^j(p_T,y)=\frac{\sigma_{MB}(p_T,y)}{\sigma_N(p_T,y)+\sigma_{MB}(p_T,y)}\,.
\label{eq:100}
\eeq
In our model in jet events soft interaction of the projectile proton
and the nucleus with the probability $W_N^j$ occurs as $N+A$ collision and 
with the probability $W_{MB}^j$ as $MB+A$ collision. 

 As usual, we define the 
minimum bias centrality $c$ through the theoretical charged multiplicity 
distribution $P$ \cite{Broniowski_c}
\beq
c(N_{ch})=\sum_{N=N_{ch}}^{\infty}P(N)\,.
\label{eq:110}
\eeq
Here $N_{ch}$ is the theoretical charged multiplicity in the pseudorapidity
window used for the centrality categorization (as in \cite{Z_MCGL1,Z_MCGL2}
we use the pseudorapidity region $|\eta|<0.5$).
It is important that the number of the binary collisions 
in the denominator of (\ref{eq:10}) is defined in terms of the centrality
classes for the minimum bias events. 
The centrality dependence of $R_{pA}$ arises due to the fact that
the shapes of the charged multiplicity distributions for the minimum bias 
soft events (used in the centrality selection) and for the jet 
events are different. 
For a given centrality class $\{c\}$ $N_{coll}$ can be written as \cite{Steinberg}
\beq
N_{coll}(\{c\})=\frac{\sigma_{in}^{NN}}{\sigma_{in}^{pA}}
\int d\bb T(b)P_s(\{c\},b)\,,
\label{eq:120}
\eeq
where $P_s$ is the probability that multiplicity of the UE belongs 
to the centrality class $\{c\}$ and $T$ is the impact parameter 
probability distribution of 
the binary collisions. In the approximation of zero interaction radius
$T$ is reduced to the nuclear profile function $T_A(\bb)=
\int dz\rho_A(\bb,z)$ ($\rho_A$ is the nuclear density).
In our two-component model $P_s$ can be written as
\beq
P_s(\{c\},b)=W_NP_{N}(\{c\},b)+W_{MB}P_{MB}(\{c\},b)\,,
\label{eq:130}
\eeq 
where $P_{N,MB}(\{c\},b)$ are the centrality probabilities for 
the $N+A$ and $MB+A$ collisions.
In calculation of the numerator of (\ref{eq:10}) the probability that in
a jet event the multiplicity for the UE belongs to the centrality class
$\{c\}$ (we denote it $P_j$) can be written via $W_{N}^j$ and
$W_{MB}^j$ as 
\beq
P_j(\{c\},b,p_T,y)=W_N^{j}(p_T,y)P_{N}(\{c\},b)+W_{MB}^j(p_T,y)P_{MB}(\{c\},b)\,.
\label{eq:140}
\eeq
In terms of the probabilities (\ref{eq:130}), (\ref{eq:140}) the theoretical 
$R_{pA}$ can be written as
\beq
R_{pA}(\{c\},p_T,y)=\frac{R(p_T,y)\int d\bb T(b)P_j(\{c\},b,p_T,y)}
{\int d\bb T(b)P_s(\{c\},b)}\,.
\label{eq:150}
\eeq
Here the factor $R$ account for modification of the hard cross section
due to the nuclear modification of the PDFs of bound nucleons (in nucleus).
For the whole centrality range
$\{c\}=(0,1)$ $P_{j,s}=1$ and $R_{pA}$ is simply reduced to $R$.

\section{Numerical results and discussion}
In Fig.~1 we show $f_{M/p}$ splitting function used in (\ref{eq:40})
obtained with $\gamma_5$ pion-nucleon vertex for the dipole formfactor 
(\ref{eq:70}). 
\begin{figure}[ht]
\epsfig{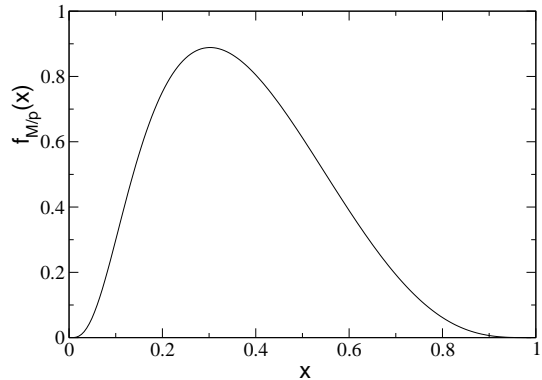}
\caption{\small 
$f_{M/p}$ splitting function normalized to $W_{MB}=0.4$  obtained  
using the $x$-distribution for the $\pi N$ Fock component. 
}
\end{figure}
One can see that the meson spectrum is strongly peaked near $x\sim 0.3$.
For the bare meson PDFs in (\ref{eq:40}) we use the LO parametrization 
of the pion PDFs of \cite{pionPDF}. 
For the bare nucleon PDFs we use the LO CTEQ6 \cite{CTEQ6} parametrizations. 
For the PDF momentum scale $Q$ for pion we use the hard parton transverse 
momentum $p_{T}$. In our model, due to the presence of the $MB$ component,
the transverse mean-square radius of the bare proton becomes smaller 
by a factor of $a\approx 0.88$. To account for a possible decrease of the 
range of the DGLAP evolution due to a bigger initial momentum scale 
we take $Q=a p_T$ for the bare nucleon PDFs.
However, the choice $Q=p_T$ gives practically same results. 
For the nucleons in a lead nucleus we use the LO CTEQ6 \cite{CTEQ6} PDFs with
the EKS98 \cite{EKS98} nuclear corrections.
The hard cross sections have been calculated using the LO 
pQCD formula. To simulate the higher order effects 
for the virtuality scale in $\alpha_{s}$ we take the value 
$cQ$ with $c=0.265$ as in the PYTHIA event generator \cite{PYTHIA}.
This gives a fairly good description 
of the $p_{T}$-dependence of the inclusive jet cross section obtained 
in \cite{ATLAS1}.
\begin{figure}[ht]
\epsfig{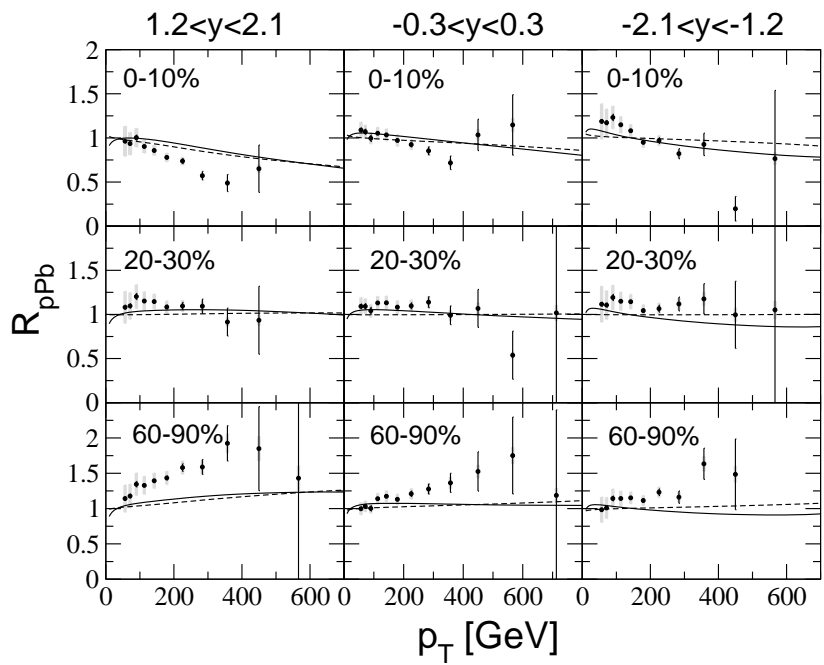}
\caption{\small $R_{pPb}$ versus $p_T$ for $p+Pb$ 
collisions at $\sqrt{s}=5.02$ TeV
for 0-10\% (upper), 20-30\% (middle), $60-90$\% (lower) centrality classes.
The solid curves show our results with the EKS98 correction factor $R$
in (\ref{eq:150}), and the dashed ones without $R$.
Data points are from ATLAS \cite{ATLAS1}.
}
\end{figure}

We have computed the numerator and denominator
of (\ref{eq:150}) by sampling $pA$ collisions within
the MCG model with meson cloud developed in \cite{Z_MCGL1,Z_MCGL2}.
The interested reader is referred to those papers for details of our 
MCG scheme.
In Fig.~2 we compare our results
with the ATLAS data \cite{ATLAS1}. To illustrate the effect of the
nuclear modification of the nucleon PDFs, in Fig.~2 we present the 
results obtained with and without the EKS98 correction factor $R$
in (\ref{eq:150}).
From Fig.~2 one sees that the effect of the meson cloud on the 
$R_{pA}$ shows qualitatively the same tendency as that observed 
by ATLAS \cite{ATLAS1}. 
The $MB$ component suppresses the central jet events 
and enhances the peripheral jet events.
Similarly to the ATLAS data the effect is more pronounced  at $y>0$.
However, quantitatively the effect is somewhat smaller than in the
data. It is possible that a better agreement 
with the ATLAS data \cite{ATLAS1} can be obtained by accounting 
for the correlations between hard and soft partons in 
the bare constituents in the projectile proton, due to the mechanism 
discussed in \cite{Skokov1,Armesto,Majumder}.
We postpone this for future work.
Also, it is possible that the meson effects in the proton PDFs 
can be enhanced due to the 
nonperturbative 
quark-gluon-pion anomalous chromomagnetic interaction 
related to the instantons discussed 
recently in \cite{Kochelev}.

\begin{acknowledgments} 	
I thank N.I.~Kochelev for discussion of the results of \cite{Kochelev}.
This work is supported 
in part by the 
grant RFBR
15-02-00668-a.
\end{acknowledgments}


\section*{References}
\begin{thebibliography}{99}

\bibitem{Collins1}
J.C. Collins, D.E. Soper, and G.F. Sterman,
Adv. Ser. Direct. High Energy Phys. {\bf 5}, 1 (1989)
[hep-ph/0409313].

\bibitem{ATLAS1}
G. Aad {\it et al.}  [ATLAS Collaboration],
Phys. Lett. B{\bf 748}, 392 (2015) 
[arXiv:1412.4092].



\bibitem{Z_mQGP2}
B.G. Zakharov,
J. Phys. {\bf G41}, 075008  (2014) 
[arXiv:1311.1159]; AIP Conf.~Proc. {\bf 1701}, 060001 (2016) 
[arXiv:1412.0295].



\bibitem{Z_mQGP1}
B.G. Zakharov,
Phys. Rev. Lett. {\bf 112}, 032301 (2014)
[arXiv:1307.3674].


\bibitem{Skokov1}
A. Bzdak, V. Skokov, and S. Bathe,
Phys. Rev. C{\bf 93}, 044901  (2016)
[arXiv:1408.3156].

\bibitem{Armesto}
N. Armesto, D.C. G\"ulhan, and J.G. Milhano,
Phys. Lett. B{\bf 747}, 411 (2015)
[arXiv:1502.02986].

\bibitem{Majumder}
M. Kordell and A. Majumder, arXiv:1601.02595.

\bibitem{ST}
J. Speth and  A.W. Thomas,
Adv. Nucl. Phys. {\bf 24}, 83 (1997).

\bibitem{Z_MCGL1}
B.G. Zakharov,
JETP Lett. {\bf 104}, 6 (2016) 
[arXiv:1605.06012].

\bibitem{Z_MCGL2}
B.G. Zakharov, JETP in press [arXiv:1611.05825].

\bibitem{Zoller_MB}
V.R. Zoller,
Z. Phys. C{\bf 60}, 141 (1993).


\bibitem{Zoller_MB1}
V.R. Zoller,
Z. Phys. C{\bf 53}, 443 (1992).

\bibitem{MST_MB}
W. Melnitchouk, J. Speth, and A.W. Thomas,
Phys. Rev. D{\bf 59}, 014033 (1998) 
[hep-ph/9806255].

\bibitem{Broniowski_c}
W. Broniowski and W. Florkowski,
Phys. Rev. C{\bf 65}, 024905 (2002) 
[nucl-th/0110020].

\bibitem{Steinberg}
M.L. Miller, K. Reygers, S.J. Sanders, and P. Steinberg,
Ann.~Rev.~Nucl.~Part.~Sci. {\bf 57}, 205 (2007)
[nucl-ex/0701025].


\bibitem{CTEQ6}
S.~Kretzer, H.L.~Lai, F.~Olness, and W.K.~Tung,
Phys. Rev. D{\bf 69}, 114005 (2004).


\bibitem{pionPDF}
M. Gluck, E. Reya, and I. Schienbein,
Eur. Phys. J. C{\bf 10}, 313 (1999)
[hep-ph/9903288].

\bibitem{EKS98}
K.J.~Eskola, V.J.~Kolhinen, and C.A.~Salgado,
Eur. Phys. J. C{\bf 9}, 61 (1999).


\bibitem{PYTHIA}
T.~Sjostrand, L.~Lonnblad, S.~Mrenna, and  P.~Skands,
arXiv:hep-ph/0308153.

\bibitem{Kochelev}
N. Kochelev, H.-J. Lee, B. Zhang, and P. Zhang,
Phys.~Lett. B{\bf 757}, 420 (2016)
[arXiv:1512.03863].


\end{thebibliography}
\end{document}